\documentstyle[aps,epsf]{revtex}

\def\be{\begin{equation}}
\def\ee{\end{equation}}
\def\bea{\begin{eqnarray}}
\def\eea{\end{eqnarray}}

\begin{document}
 \newcommand{\h}{\mbox{\sf h}}
 \newcommand{\hk}{\mbox{\footnotesize \sf h}}
 \newcommand{\ew}{\mbox{\tiny EW}}
 \newcommand{\kpz}{\mbox{\tiny KPZ}}

\title{
{\small{Contribution to {\it Chaos and Fractals in Chemical Engineering},
Rome 1996 }}\vspace{1cm}\\
DOMAIN WALL ROUGHENING IN DISORDERED MEDIA:\\
FROM LOCAL SPIN DYNAMICS TO A\\
CONTINUUM DESCRIPTION OF THE INTERFACE
}

\author{M. Jost \raisebox{0.5ex}{*}, K.~D. Usadel \raisebox{1ex}{$\diamond$} }

\address{
Theoretische Tieftemperaturphysik, 
Gerhard-Mercator-Universit\"at Duisburg,\\ 
Lotharstr.~1, 47048 Duisburg, Germany \\
}

\maketitle

\begin{abstract}
We study the kinetic roughening of a driven domain wall between 
spin-up and spin-down domains for
a model with non-conserved order parameter and quenched disorder.
To understand the scaling behavior of this interface we construct
an equation of motion and study it theoretically. 
\end{abstract}

\section{Introduction}
Roughening phenomena of growing surfaces and moving interfaces are of great interest
for more than one decade since they appear in a variety of technological and
scientific problems. The most interesting topic in the field of surface growth
is the molecular beam epitaxy (MBE) due to its technical importance~\cite{Wolf}.
Examples for the roughening of moving interfaces are the immiscible-fluid 
displacement in oil recovery and the dynamics of magnetic domain walls, see for 
example~\cite{KopLev}. Typical for the interface problem is the occurrence of 
the so-called depinning transition~\cite{Kardar}. Depending on a driving 
force the interface gets trapped by random impurities, i.e. it is in the 
{\it pinning} phase. For values
above the critical force the interface moves steadily with non-zero 
velocity, it is in the {\it depinning} phase.
In the examples listed these phenomena occur on very
different length scales.
For a description of all of these phenomena the concepts of self-affine
fractals and dynamic scaling are of crucial importance~\cite{Fam}.

Generally it is supposed that the surfaces as well as the interfaces can be 
described by an 
equation of motion (EOM) which belongs to one of the famous universality classes,
namely the Edwards-Wilkinson (EW)~\cite{EW} or the 
Kardar-Parisi-Zhang (KPZ)~\cite{KPZ} universality class.
In the first case the EOM for the surface (interface) profile
function $\h(x,t)$ reads 
\be
 \frac{\partial \h}{\partial t} = \nu \nabla^2 \h + \eta +F 
\label{gew}
\ee
with a noise term $\eta$ and the driving force $F$. In the second case the so-called
KPZ nonlinearity $\frac{\lambda}{2} (\nabla \h)^2$ with $\lambda$ proportional to 
the surface (interface) velocity occurs additionally in Eq.~(\ref{gew}).
In the case of growing surfaces the noise arises mainly due to 
statistical fluctuations of the deposition rates and thus it can 
be assumed that the noise is time dependent (annealed). In contrast the noise
in the other examples listed arises from the moving of the interface through 
a random background, represented by a time independent (quenched) noise term.

In the present article we study the motion and the morphology of a domain wall in a 
ferromagnetic medium containing quenched random fields.
A continuum description of the interface is given and possible equations 
of motion are discussed.

\section{Model}
To study the behavior of a domain wall in a ferromagnetic medium (low temperature phase) 
we start from a Ginzburg-Landau type Hamiltonian
\be
  {\cal{H}} =  \int d^d{\bf r}\, 
  \Bigl(-\frac{a}{2}\phi({\bf r},t)^2+\frac{b}{4}\phi({\bf r},t)^4
  +\frac{J}{2} \left(\nabla \phi({\bf r},t)\right)^2
  - (H+B({\bf r}))\,\phi({\bf r},t) \Bigr)
\label{gl}
\ee
where $\phi$ denotes a scalar order parameter, $H$ denotes a homogeneous
driving field and $B({\bf r})$ a quenched random field.
This continuum description of a ferromagnetic material on a mesoscopic
scale is well established.
It is assumed that $a=b=u_0$ without loss of generality 
and that the random fields, drawn with 
equal probability, have zero mean and are uncorrelated in space.
Furthermore we assume $J>0$ corresponding
to ferromagnetic spin-spin coupling and $u_0>0$ for stability.
The dynamics of the system for non-conserved
order parameter is defined through a Langevin equation
\be
\gamma \frac{\partial \phi({\bf r},t)}{\partial t}
 = - \frac{\partial {\cal{H}}}{\partial \phi({\bf r},t)}
\label{tdgl}
\ee
with a relaxation time proportional to $\gamma$
(model A in the classification of Hohenberg and Halperin~\cite{HohHal}), for further 
details see \cite{mjt_pre,usa}.
Temperature can be neglected since from renormalization group studies 
it is believed to be irrelevant~\cite{zero}. Another argument for the
irrelevance of temperature has been given by Grinstein and Ma~\cite{grin} for 
the random field Ising model (RFIM) in $d$ dimensions 
using scaling arguments for the interface 
roughness $w$ (defined in the next section). 
They found that it scales for temperature $T=0$
as $w^2 \sim L^{2(4-{\cal D})/3}$ with domain size $L$ in the interface dimension 
${\cal D}=d-1$.
In the pure Ising model it scales as $L^{2-{\cal D}}$ for $T>0$. Since $2(4-{\cal D})/3$
is for all ${\cal D}$ greater than $2-{\cal D}$ the field-induced perturbations dominate
the thermal perturbations and thus temperature can be neglected.
This argument can be applied to the present model since for 
$a \rightarrow \infty$, $b \rightarrow \infty$ and $a/b=1$ one recovers
the RFIM~\cite{bin}.

\section{Dynamic scaling of moving interfaces}
We assume a geometry where a horizontal domain wall separates one ferromagnetic domain
with positive magnetization below the wall and one with negative magnetization
above the wall.
The position $y=\h({\bf x},t)$ of the interface is then defined as that point
${\bf r}=({\bf x},y)$ at which 
$\phi({\bf x},y,t)$ as function of $y$ for fixed ${\bf x}$ changes sign.
Numerical studies~\cite{mjt_pre} starting from a discretized version of Eq.~(\ref{tdgl})
showed that 
$\h({\bf x},t)$ is a single valued function for not too large driving and random fields. 

Starting from an initially flat interface
the interface develops a rough structure due to the random fields which increase
or decrease the driving force locally. This roughness can be measured by the 
root-mean-square fluctuations of the averaged interface position~\cite{Fam}
\be
w(L,t)=\left\langle \left(\h({\bf x},t) 
- \langle \h({\bf x},t) \rangle\right)^2 \right\rangle^{1/2} 
\label{width}
\ee
which is usually called the roughness. Here $L$ denotes the system size perpendicular
to the moving direction of the interface and the angular brackets denote
an average over all lattice sites at positions ${\bf x}$ as well as over different
realizations of the disorder. Another quantity of interest is 
the height correlation function 
\be
  C(r,t)=\langle [\h({\bf x}+{\bf r},t)-\h({\bf x},t)]^2 \rangle \, ,
\label{hcf}
\ee
which is related to the roughness $w(L,t)$ according to
\be
 w(L,t)^2= \frac{1}{2L^{\cal D}} \sum_{{\bf r}} C(r,t) \, .
\label{wh}
\ee
Eq.~(\ref{wh}) is exact for periodic boundary conditions and for open boundary
conditions for $L \rightarrow \infty$.
We will concentrate in the following on the scaling behavior of the height 
correlation function as a characteristic function describing the behavior of 
the interface.

Due to the interaction terms in the underlying model the single interface elements 
do not grow independently of their neighbors and it is assumed~\cite{Fam}
that there exist a time dependent correlation length $\xi(t)$ parallel to the
interface direction.
For $r \ll \xi(t)$ the correlations of the interface do not grow further and their 
saturation value is assumed to scale like
\be 
C(r \ll \xi(t),t) \propto r^{2\alpha}
\label{hcf_a}
\ee
with the roughness exponent $\alpha$. For $r$ larger than the time dependent 
correlation length $\xi(t)$ the interface fluctuations are uncorrelated in 
space but they are still growing with time, their time evolution  can be 
described by
\be
C(r \gg \xi(t),t) \propto t^{2\beta} \,.
\label{hcf_b}
\ee
These scaling relations (\ref{hcf_a}) and (\ref{hcf_b}) are considered as 
limiting cases of a general dynamic scaling relation
\be
C(r,t)= \left(\xi(t)\right)^{2\alpha} g\left(\frac{r}{\xi(t)}\right) \, .
\label{scal}
\ee
In order that Eqs. (\ref{hcf_a}) and (\ref{hcf_b}) are recovered the scaling 
function $g(y)$ must have the properties 
$g(y) \approx \mbox{const}$ for $y\gg 1$ and $g(y) \propto y^{2\alpha}$ for $y \ll 1$.
Furthermore the correlation length has to grow with time according to 
\be
\xi(t) \propto t^{1/z}
\label{xit}
\ee
with the dynamic exponent $z=\alpha/\beta$. At time $\tau \sim L^z$ $\xi(t)$ is 
of the order of the system size and it cannot grow further thus a crossover 
to a finite size behavior occurs. 

From these scaling assumptions
follows that the interface scales as 
\be
 \tilde{\h}({\bf x},t) \sim t^{\beta}f\left(\frac{x}{t^{1/z}}\right)
\label{hdyn}
\ee
showing that the interface can be conceived as a statistical self-affine fractal
on length scales $x \ll \xi(t)$.
Here $\tilde{\h}({\bf x},t)=\h({\bf x},t)-\langle \h({\bf x},t)\rangle$ denotes 
the deviation of the interface from its time dependent average and $f(\rho)$ denotes
a scaling function with the limiting properties 
$|f(\rho)| \sim const$ for $\rho \gg 1$ and $f(\rho) \sim \rho^{\alpha}$ for 
$\rho \rightarrow 0$.

\section{Continuum description of the interface}
We start the discussion of Eq.~(\ref{tdgl}) by considering first the 
case $B({\bf r})=0$. 
The corresponding Langevin equation (\ref{tdgl}) reads
\be
\frac{\partial \phi({\bf r},t)}{\partial t} =
 -u_0 \phi({\bf r},t)\left(\phi({\bf r},t)^2-1\right)
 +J \bigtriangleup \phi({\bf r},t) +H  \, .
\label{bzero}
\ee
Due to a driving field $H>0$ the domain wall moves in the positive $y$-direction and
an approximate ansatz for the solution of Eq.~(\ref{tdgl}) is
\be
\phi({\bf r},t)=\psi\left(y+vt\right) 
\label{ab0}
\ee
where translational invariance perpendicular to the $y$-direction is assumed.
The velocity $v$ of the domain wall is given by~\cite{langer,phd_l} 
\be
 v \simeq \frac{3}{2}H\Omega_0 
\label{v}
\ee
where $\Omega_0=\sqrt{2J/u_0}$ is the intrinsic width normal to the domain wall
in the static problem $ \frac{\partial \phi({\bf r},t)}{\partial t}\equiv 0$ ($H=0$).

Taking local curvature elements of the domain wall into account an ansatz 
\be
 \phi({\bf r},t)= \psi\left(\frac{y+\h({\bf x},t)}{\Omega}\right)
\label{ansatz}
\ee
where $\h({\bf x},t)$ denotes the time dependent wall position and
\be
\Omega=\Omega_0\sqrt{1+(\nabla \h)^2} 
\label{om}
\ee
the local width of the wall in $y$-direction has been assumed~\cite{zia}.
Within this ansatz a KPZ-like EOM
\be
 \frac{\partial \h}{\partial t}=\nu\nabla^2 \h
 +\Lambda\frac{\partial \h}{\partial t} \left(\nabla \h \right)^2
 + \eta\left({\bf x},\h({\bf x},t)\right) +F
\label{eom_a}
\ee
with $\nu=\nu(J)$, effective magnetic fields $\eta$ and $F$ and the noise correlator
\be
\langle\eta({\bf x}_0,\h_0)\eta({\bf x}_0+{\bf x},\h_0+\h)\rangle
 = D \delta^{\cal D}({\bf x}) \delta(\h) 
\label{eomc}
\ee
can be obtained for a medium with random fields, see Ref.~(14) for a 
similar approach. For an analysis of the EOM it is reasonable to substitute the 
time derivative $\partial \h /\partial t$ on the right side of Eq.~(\ref{eom_a})
by the averaged interface velocity $v$ so that Eq.~(\ref{eom_a}) can be replaced by
\be
 \frac{\partial \h}{\partial t}=\nu\nabla^2 \h
 +\frac{\lambda}{2} \left(\nabla \h \right)^2
 + \eta\left({\bf x},\h({\bf x},t)\right) +F
\label{eom}
\ee
where $\lambda/2$ denotes the product of $\Lambda$ and $v$.

\section{Discussion of the EOM}
In the quenched disorder case considered here the exponents 
characterizing the scaling behavior are not known exactly. 
We will show that with special assumptions  at the depinning 
transition, $F \simeq F_C$, and for the fast moving interfaces, $F \gg F_C$, 
the values of the exponents can be obtained.

At the depinning transition the average velocity of the domain wall is
zero. Therefore Eq.~(\ref{eom}) is reduced to the quenched form of the EW equation
\be
 \frac{\partial \h}{\partial t}=\nu\nabla^2 \h 
+ \eta\left({\bf x},\h({\bf x},t)\right) +F \, .
\label{qew}                   
\ee
The average over the noise at the pinned location of the interface,
$\langle\eta({\bf x},\h({\bf x},t))\rangle$, compensates the driving force $F$ 
at the depinning transition so that only the noise fluctuations
$\delta\eta({\bf x},\h({\bf x},t))=\eta({\bf x},\h({\bf x},t))
-\langle\eta({\bf x},\h({\bf x},t))\rangle$
are relevant in the EOM which then reads
\be
 \frac{\partial \h}{\partial t}=\nu\nabla^2 \h 
+ \delta\eta\left({\bf x},\h({\bf x},t)\right)   \, .
\label{qew2}                   
\ee
For the correlations of the noise fluctuations we again assume
\be
\langle\delta\eta({\bf x}_0,\h_0)\delta\eta({\bf x}_0+{\bf x},\h_0+\h)\rangle
 = D \delta^{\cal D}({\bf x}) \delta(\h) \, .
\label{qewc1}
\ee

Describing $\h$ as a generalized self-affine function with time dependent correlation 
length, as discussed above, a rescaling transformation
\be
\hat{\h}({\bf x},t)=a^{-\alpha}\h(a{\bf x},a^z t)
\label{scale_h}
\ee
with a scaling factor $a>1$ can be used to obtain the roughness exponent $\alpha$
and the dynamic exponent $z$. Under such a scaling transformation the 
noise, according to  Eq.~(\ref{qewc1}) scales as 
\be
\delta\hat{\eta}({\bf x},\h)=a^{({\cal D}+\alpha)/2}\delta\eta(a{\bf x},a^{\alpha} \h) \, .
\label{scale_eta}
\ee
Inserting now Eq.~(\ref{scale_h}) and Eq.~(\ref{scale_eta}) into  
Eq.~(\ref{qew2}) the EOM is recovered with rescaled coefficients
\be
\hat{\nu}=a^{z-2} \nu\, ,\mbox{\hspace{0.3cm}}
\hat{D}=a^{2z-{\cal D}-3\alpha} D\, .
\label{res_co}
\ee
With $z=2$ and $\alpha=(2z-{\cal D})/3$ the coefficients $\nu$ and $D$ remain 
invariant and this invariance of the coefficients determines the values of the exponents. 
Therefore, the small time exponent $\beta$ has the value 
$\beta=\alpha/z=(2z-{\cal D})/3z$. 
In simulations of the spin model~\cite{mjt_pre,mjt_3d} Eq.~(\ref{tdgl}) we found near 
the depinning transition $\alpha \approx 0.9$, $z \approx 2$ in ${\cal D}=1$ and 
$\alpha \approx 0.68$, $z \approx 1.89$ in ${\cal D}=2$
in good agreement with our theoretical values obtained from the scaling analysis.

Our exponent relation for the roughness exponent $\alpha$ 
is in agreement with the prediction of Grinstein and 
Ma~\cite{grin} who found for the RFIM $\alpha=(4-{\cal D})/3$ (see above).
On the other hand, Leschorn~\cite{lesch} found for an 
automaton model of Eq.~(\ref{qew}) in the interface dimension ${\cal D}=1$ the 
values $\alpha \approx 1.25$ and $\beta \approx 0.88$ and
$\alpha \approx 0.75$ and $\beta \approx 0.48$ in ${\cal D}=2$. 
Renormalization group (RG) 
studies of Eq.~(\ref{qew}) by Nattermann et al.~\cite{natter} in ${\cal D}=4-\epsilon$
lead to $\alpha=\epsilon/3$ in agreement with our calculation 
and $z=2-2\epsilon/9$ which disagrees with our ${\cal D}$-independent value for 
the dynamic exponent. Leschhorn notes that the
dynamic exponent $z$ obtained  numerically from a measurement of $\alpha$ and $\beta$
through $z=\alpha /\beta$ is very close to the RG values for different ${\cal D}$. 
With dimensional analysis methods Parisi~\cite{parisi_les} also found 
$\beta=(4-{\cal D})/6$ but this relation was questioned by him due to numerical 
results of an EOM similar to Eq.~({\ref{qew}) presented in a
previous paper~\cite{parisi}.
He also gave arguments in favor of the scaling relation $\beta=(4-{\cal D})/4$
in Ref.~(19).
In contrast, Kessler et al.~\cite{kessler} as well as  Dong et al.~\cite{dong}
found numerically $\alpha \approx 1$ for ${\cal D}=1$ 
and Ji et al.~\cite{ji} $\alpha \approx 0.67$ for ${\cal D}=2$ 
close to our values. However, discrepancies
between the results of Dong et al. and of Ji et al. 
and our results still exist. Measuring the
velocity dependence of $F-F_C$ Dong et al. found $v \propto (F-F_C)^{\theta}$ with 
$\theta \approx 0.25$ while we found earlier~\cite{usa} $\theta \approx 1$.
Ji et al. found no self-affine scaling in ${\cal D}=1$.

Since our results, both numerically and theoretically, for the roughness exponent
$\alpha$ agree with other numerical work but disagree for the small time exponent 
$\beta$ we also integrated numerically the EOM Eq.~({\ref{qew}) in ${\cal D}=1$
in order to understand this discrepancy.
We integrate a system of size $L=4096$ up to a time $t \approx3.3 \, 10^5$
within an Euler-scheme with $dt=0.02$ and as the lattice spacing in $\h$-direction we 
chose $d\h=0.001$. For the noise $\eta$ uniformly distributed in $[-0.2,0.2]$
we observed a depinning transition close to $F=0.012$. 
The results presented here are obtained for this driving force.
Fig.~(1a) shows the spatial correlations of the height correlation function
$C(r,t)$ as function of the distance $r$ for different times on logarithmic scales.
An $r$-independent saturation is observed for large $r$ in agreement with 
Eq.~(\ref{hcf_b}) and a linear dependence on $\ln(r)$ for small $r$ in agreement
with Eq.~(\ref{hcf_a}). For the corresponding exponent
we found $\alpha=0.98$ in good agreement with our theoretical result.
However, the crucial point is that $\ln(C(r,t))$ is still $t$-dependent for small $r$,
i.e. contrary to the scaling prediction Eq.~(\ref{hcf_a}) the correlations are still
growing with time, $C(1,t) \propto t^{2\kappa}$ with $\kappa=0.2$
(see Fig.~(1b)). This was also found previously in simulations of the corresponding 
spin model~\cite{mjt_pre} Eq.~(\ref{tdgl}). Note that
such a behavior can be found in many different roughening models~\cite{siegert}.
One possibility to take this anomalous behavior into account is a modification of 
the scaling relation Eq.~(\ref{scal}) in the following way~\cite{mjt_pre} (see Fig.~(1c) 
for a scaling plot):
\be
C(r,t)=\xi(t)^{2\alpha/\lambda} g\left(\frac{r}{\xi(t)}\right) \, ,
\label{sc2}
\ee
with the limiting forms 
\be
C(r\gg \xi(t),t) \propto \xi(t)^{2\alpha/\lambda}
\label{n_beta}
\ee
and
\be
C(r\ll \xi(t),t) \propto r^{2\alpha} \xi(t)^{2\alpha(1/\lambda-1)}
\ee
where we have set $\lambda=1-\kappa/\tilde{\beta}$ for convenience. 
Note that the exponent $\alpha$ again follows from the slope of
$\ln(C)$ versus $\ln(r)$.

\begin{figure}[h]
\begin{center}
\hspace{-0.8cm}
 \makebox[5.5cm][l]{
 \epsfxsize=5.5cm
 \epsfysize=5.5cm
 \epsffile{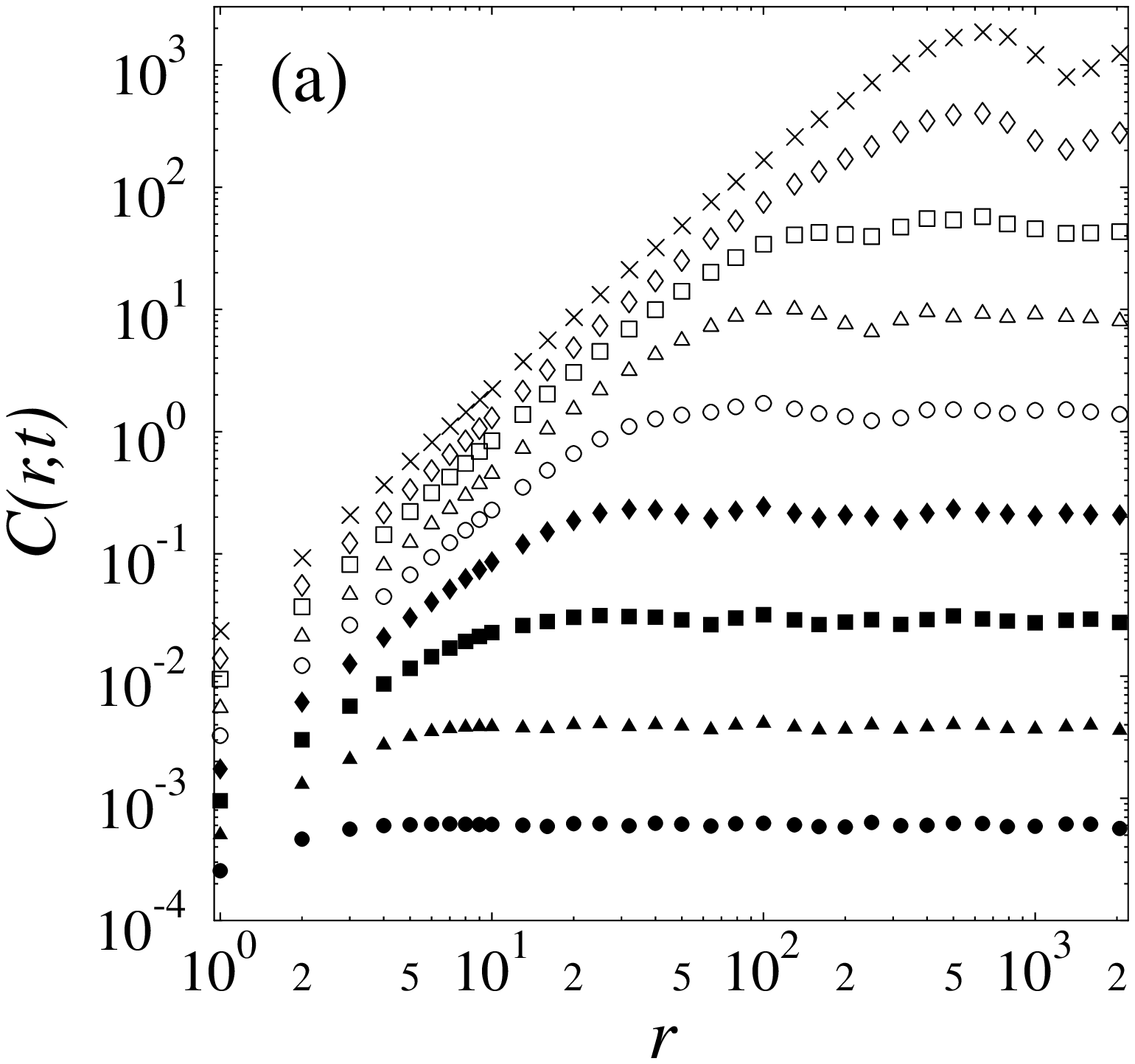} 
 }
 \makebox[5.5cm][l]{
 \epsfxsize=5.5cm
 \epsfysize=5.5cm
 \epsffile{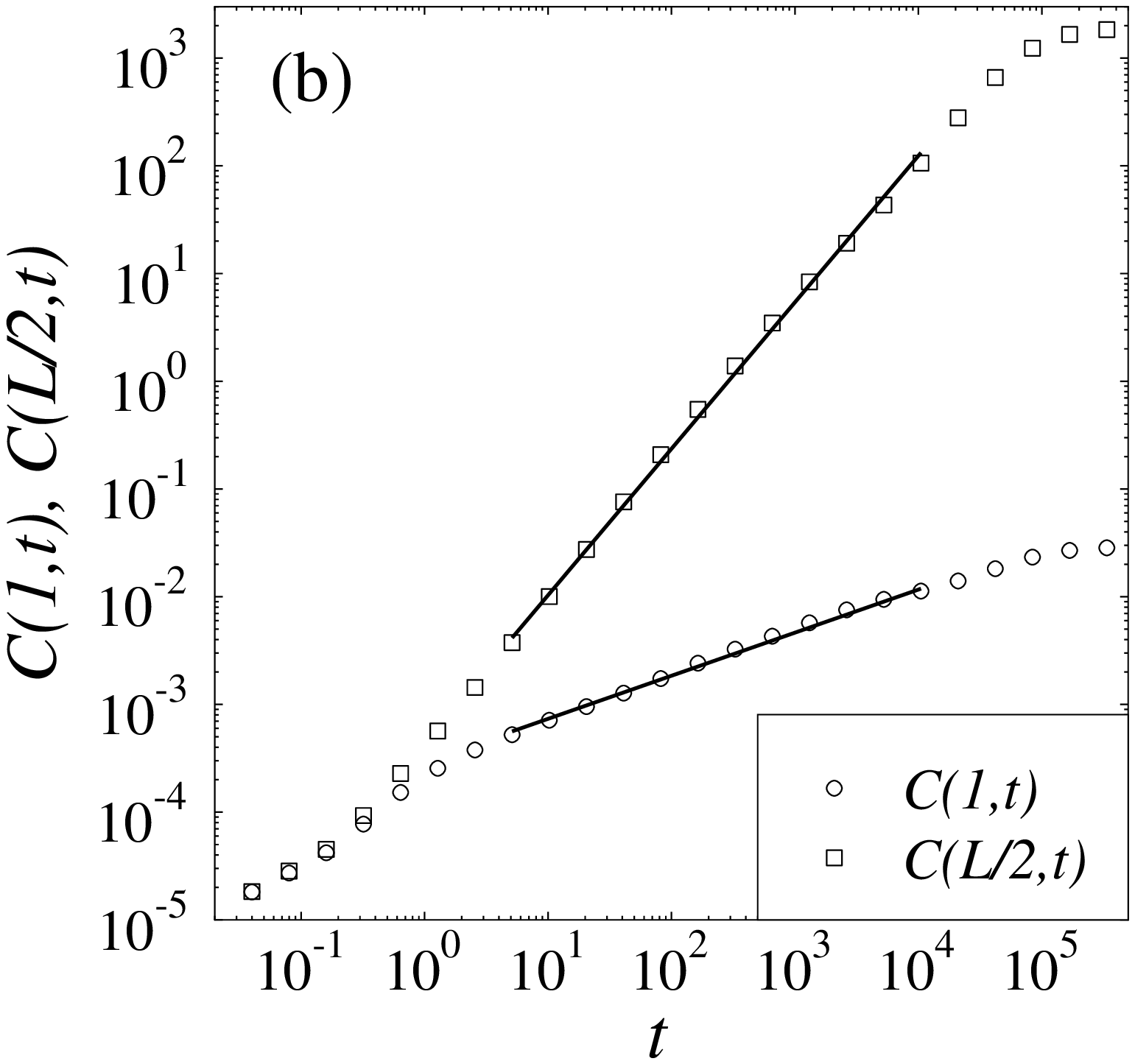} 
 }

\hspace{-0.8cm}
 \makebox[5.5cm][l]{
 \epsfxsize=5.5cm
 \epsfysize=5.5cm
 \epsffile{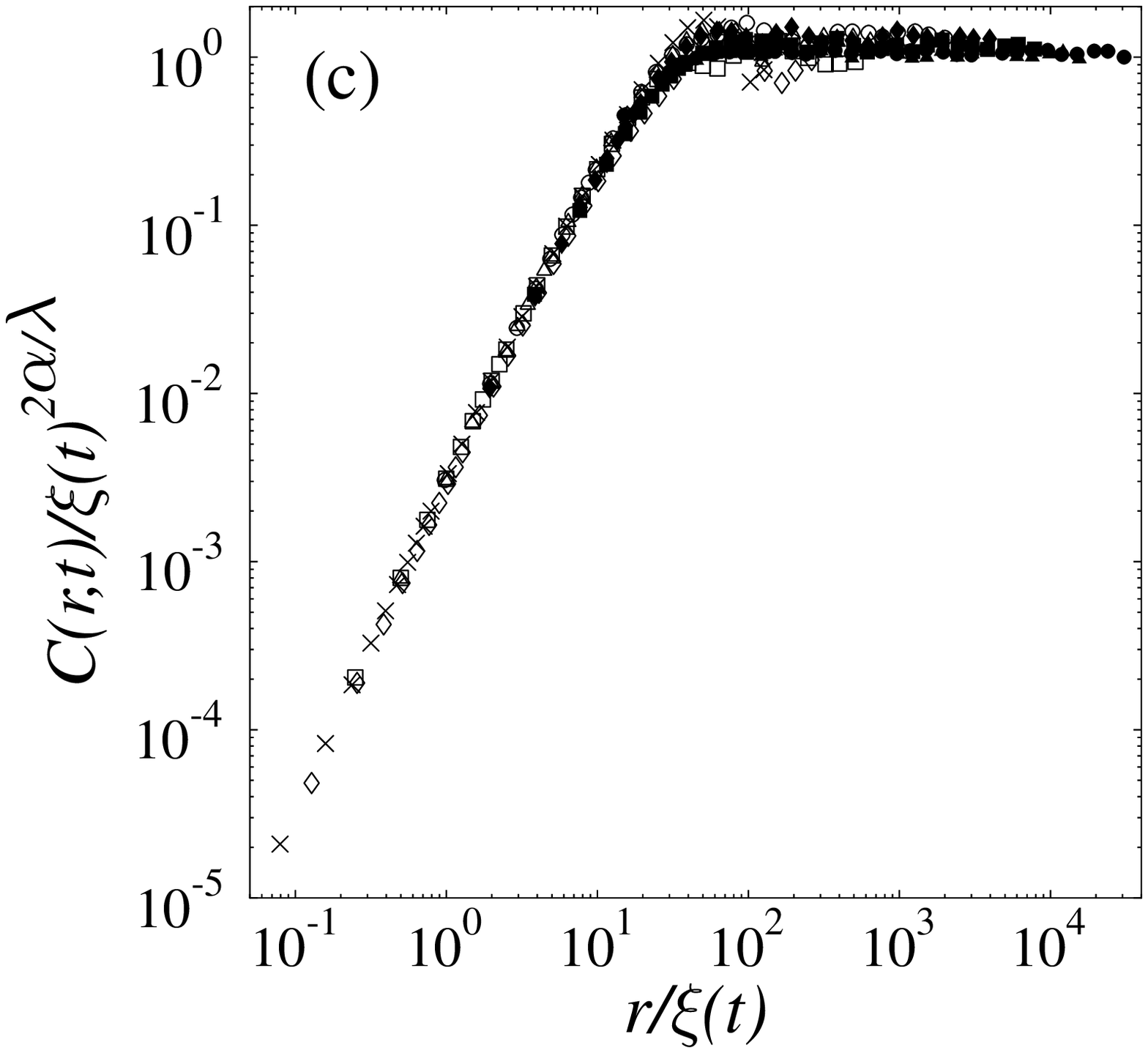} 
 }
 \makebox[5.5cm][l]{
 \epsfxsize=5.5cm
 \epsfysize=5.5cm
 \epsffile{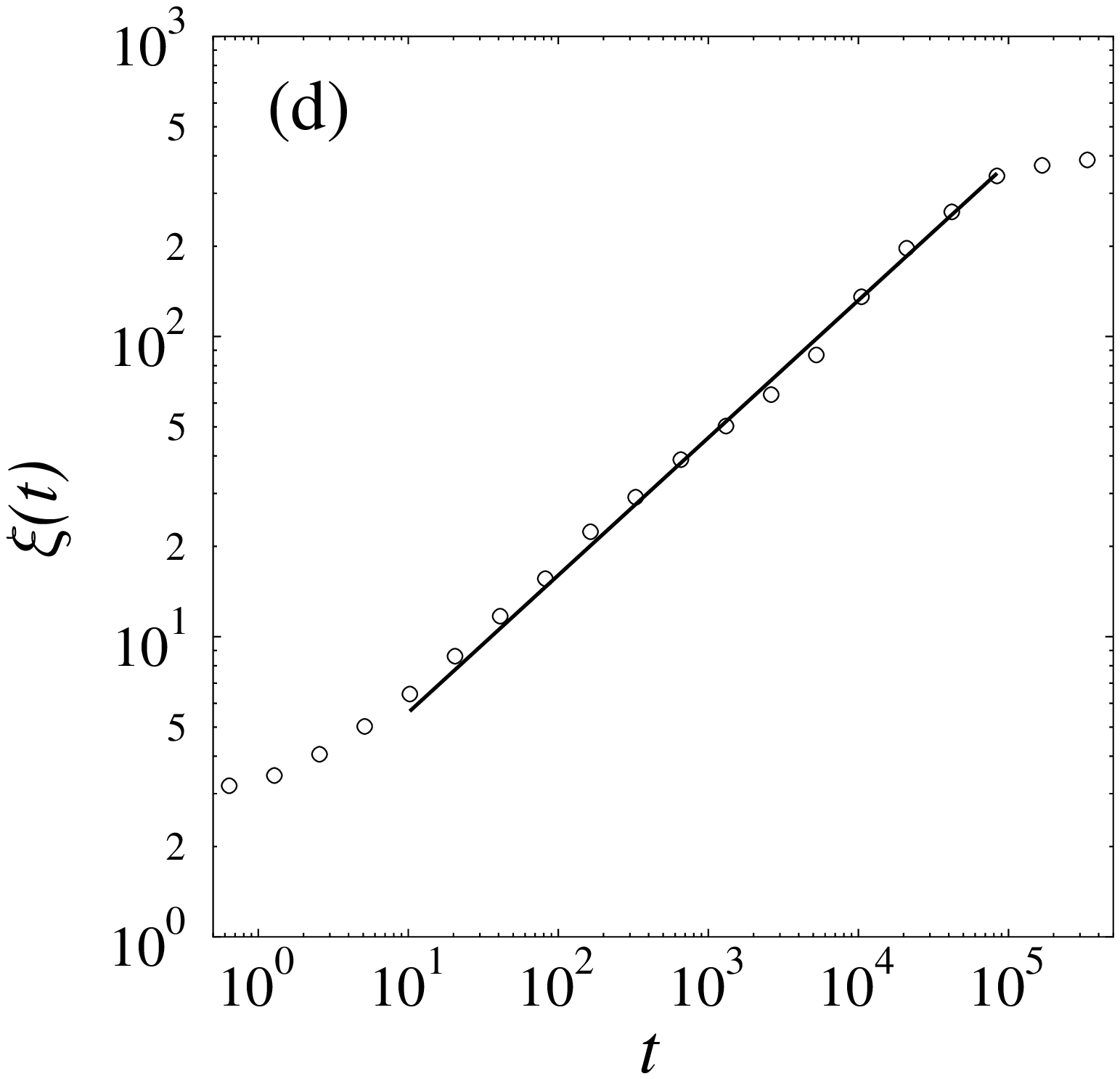} 
 }
\end{center}
\caption{
Simulation results of Eq.~(\ref{qew}) for system parameters as mentioned
in the main text. (a) The height correlation function $C(r,t)$ for different times 
$t$, $t$ increasing from bottom to top. (b) The time dependence of $C(L/2,t)$ and 
$C(1,t)$, the lines represent $C(L/2,t)\propto t^{1.36}$ and $C(1,t) \propto t^{0.4}$. 
(c) Scaling plot of the height correlation function according to Eq.~(\ref{sc2}).
(d) The time dependence of the correlation length $\xi(t)$, the line represents
a fit according to $\xi(t) \propto t^{1/z}$ with $z=2.18$.%\mbox{.\hspace{8.5cm}}$
}
\end{figure}

\noindent 
If we define the dynamical exponent $z$ as it is usual done in
dynamic scaling theory by $\xi(t) \propto t^{1/z}$ we have to set
$z=\alpha/\tilde{\beta}\lambda=\alpha/(\tilde{\beta}-\kappa)$. Therefore 
the exponent $\tilde{\beta}$ is no more 
the ratio of $\alpha$ and $z$. For the small time exponent $\tilde{\beta}$
we find from an analysis of the time dependent saturation value of the height 
correlation function at $r=L/2$ which according to Eq.~(\ref{n_beta}) can be 
written as 
\be
C(L/2,t) \propto t^{2 \tilde{\beta}}
\ee
from Fig.~(1b) as well as from the width (not shown here) $\tilde{\beta}=0.68$.
From $z=\alpha/(\tilde{\beta}-\kappa)$ we get $z=2.03$ in good agreement with
our theoretical result. 

The important result from this analysis is that only in the saturation limit in which
$C(1,t)$ is time independent, i.e. on very large time scales and therefore
for very large system sizes, the exponent $\beta$ which describes
the small time behavior of the height correlation function and of the width
is given by the ratio $\alpha/z$ and the usual scaling behavior Eq.~(\ref{scal})
is recovered. If $C(1,t)$ is still growing proportional to 
$t^{2\kappa}$ the small time behavior ($r \ll \xi(t)$) is described by the exponent 
$\tilde{\beta}=\kappa+\alpha/z$. Thus the discrepancy with the work of 
Parisi~\cite{parisi} mentioned above presumably has its origin in this
modified scaling relation since with the time dependence of the width only 
the exponent $\tilde{\beta}$ can be measured.

To further support our point of view we additionally analyze the correlation length 
$\xi(t)$ and thus the exponent $z$ directly by the following approximative method.
In a logarithmic plot of $C(r,t)$ the quantity $\xi(t)$ can be obtained 
by the intersection of a horizontal line which is defined by the time dependent 
saturation value of the height correlation function and the line which one obtains 
by fitting $C(r,t)$ for small $r$. With this method we obtain for the dynamic exponent
$z=2.18$ (see Fig.~(1d)) which is of the order of the value we obtained from the 
relation $z=\alpha/(\tilde{\beta}-\kappa)$.

The situation described so far is only valid at or very close to 
the depinning transition.
For a moving interface with larger velocity $v$ the KPZ nonlinearity 
$\lambda/2(\nabla \h)^2$ is important for the 
scaling behavior.
For a fast moving interface, $F \gg F_C$, it is plausible that
the quenched disorder acts as an effective
annealed disorder so that in this limit
the EOM is given by the well-known KPZ equation
\be
 \frac{\partial \h}{\partial t}=\nu\nabla^2 \h 
+\frac{\lambda}{2}\left(\nabla \h \right)^2 
+ \eta\left(x,t\right) 
\label{kpz}
\ee
where the constant driving force $F$ was eliminated by going in the comoving frame of 
reference and the noise has a Gaussian distribution with zero mean 
and is uncorrelated in space and time
\be
\langle \eta(x,t)\eta(x',t') \rangle = D \delta(x-x') \delta(t-t') \, .
\ee
In this case the exponents are known
exactly for ${\cal D}=1$, namely $\alpha_{\kpz}=1/2$ and $z_{\kpz}=3/2$ thus 
$\beta_{\kpz}=1/3$. In the annealed EW case these exponents are also known 
exactly and are given by $\alpha_{\ew}=1/2$ and
$\beta_{\ew}=1/4$ ($z_{\ew}=2$).
By simulation of Eq.~(\ref{kpz}) one observes a crossover at a time 
$t_C^{\ew \rightarrow \kpz} $ from an EW regime
with $\beta=\beta_{\ew}$ to $\beta=\beta_{\kpz}$ depending on the coefficients of
the KPZ equation~\cite{Krug}. Comparing the formulas for the width 
of the interface in the KPZ case,
\be
w_{\kpz}^2=c_2\left[\left(\frac{D}{2\nu}\right)^2 |\lambda| t\right]^{2/3} \, ,
\label{wkpz}
\ee
to the corresponding width in the EW case for an infinite system,
\be
w_{\ew}^2=\frac{1}{2 \pi} \frac{D}{\nu} \frac{\Gamma(1/z_{\ew})}{z_{\ew}-1} 
          \left(2\nu t\right)^{2\beta_{\ew}}
\label{wew}
\ee
one obtains for the crossover time between
an EW behavior for small times to a KPZ behavior for large times
 with the numerical estimate $c_2 \approx 0.4$ 
\be
t_C^{\ew \rightarrow \kpz} \approx 252 \nu^5 D^{-2} |\lambda|^{-4} \, .
\label{tc_ewkpz}
\ee

This relation limits the possibility of observing KPZ exponents since
$\beta$ can be observed only for times $t$ where the 
correlation length $\xi(t)$ is much smaller than the system size. For the
EW model the correlation length can be calculated exactly~\cite{Krug}
\be
 \xi(t)=(2\nu t)^{1/z_{\ew}} \,. 
\label{xietc}
\ee
Therefore, systems of length $L$ where periodic boundary conditions are assumed 
(the maximum value of the correlation length is $L/2$) saturate at times 
$t=L^{z_{\ew}}/8\nu$. 
Comparing this result with Eq.~(\ref{tc_ewkpz}) the system size $L_C$
which one needs to observe the KPZ regime has to be much larger than 
\be
  L_C \simeq 45 \frac{\nu^3}{D |\lambda|^2} \,.
\label{lmin}
\ee
We think that such a crossover from EW to KPZ behavior also
occurs in ${\cal D}=2$ where for the EW equation logarithmic dependencies
of the characteristic functions are known exactly and algebraic
behavior in the KPZ case is observed.
In our earlier simulations~\cite{mjt_pre,mjt_3d} for ${\cal D}=1$ and ${\cal D}=2$
we  observed EW behavior for all possible system sizes which 
we believe are due to these finite size effects.

\section{Conclusion}
To summarize, we argued that the motion of a domain wall in a ferromagnetic
medium described by a Ginzburg-Landau type energy can be mapped onto the quenched form
of the EW equation at the depinning transition and onto
the annealed KPZ equation for driving fields much larger than the critical field. 
In the latter case the roughness and the dynamic 
exponent are known exactly for ${\cal D}=1$ but as discussed above they
can only be observed for very large systems while for smaller systems
the annealed EW behavior is observed.
At the depinning transition we obtained the exponents from a scaling 
analysis and obtained values in agreement with our
numerical results presented earlier in ${\cal D}=1$ and ${\cal D}=2$.

%\section*{References}

\end{document}